\documentclass[journal,10pt,twocolumn,twoside]{IEEEtran}
\usepackage{multicol}   
\usepackage{multirow}
\usepackage[pdftex]{graphicx}
\usepackage{algorithm,algpseudocode}
\usepackage[small]{caption}
\usepackage{float}
\usepackage{amssymb,amsmath}
\usepackage{graphicx}
\usepackage{subfigure}
\usepackage{xspace}
\usepackage{amsthm}
\usepackage{fancyhdr}

\usepackage{caption,setspace}
\usepackage{amssymb,amsmath}
\usepackage{bm}
\usepackage{textcomp} 
\usepackage{epstopdf}
\usepackage{bbding}
\usepackage{cite} 
\usepackage{color}
\usepackage{pifont}
\usepackage{lipsum}
\usepackage{amstext}
\usepackage{orcidlink}

\usepackage{cuted}

\graphicspath{{fig/}}

\newcommand\scalemath[2]{\scalebox{#1}{\mbox{\ensuremath{\displaystyle #2}}}}

\begin{document}

\title{\huge Robust Beamforming Optimization for STAR-RIS Empowered Multi-User RSMA Under Hardware Imperfections and Channel Uncertainty}
\author{Muhammad Asif \,\orcidlink{0000-0002-9699-1675}, Asim Ihsan \,\orcidlink{0000-0001-7491-7178}, Zhu Shoujin \,\orcidlink{0000-0002-2615-2489}, Ali Ranjha \,\orcidlink{0000-0001-6663-3714}, Xingwang Li \,\orcidlink{0000-0002-0907-6517}, \IEEEmembership{Senior Member, IEEE}, Khaled M. Rabie \,\orcidlink{0000-0002-9784-3703}, \IEEEmembership{Senior Member, IEEE}, and Symeon Chatzinotas \,\orcidlink{0000-0001-5122-0001}, \IEEEmembership{Fellow, IEEE}

\thanks{This work was supported in part by the Key Research and Development Projects in Tongling City under Grant No. 20230201014; and in part by the Tongling University Talent Research Initiation Fund Project under Grant No. 2023tlxyrc17.}  	

\thanks{Muhammad Asif and Zhu Shoujin are with the School of Electrical Engineering, Tongling University, Anhui, Tongling, China (e-mails: masif@tlu.edu.cn, 2023028@tlu.edu.cn).
	
	Asim Ihsan and Khaled M. Rabie are with the Department of Computer Engineering and Center for Communication Systems and Sensing, King Fahd University of Petroleum \& Minerals, Dhahran, Saudi Arabia (e-mails: asim.ihsan@kfupm.edu.sa, k.rabie@kfupm.edu.sa).
	
	Ali Ranjha is with the École de Technologie Supérieure, Montréal, Quebec, Canada, (e-mail: ali-nawaz.ranjha.1@ens.etsmtl.ca).
	
	Xingwang Li is with the School of Physics and Electronic Information Engineering, Henan Polytechnic University, Jiaozuo, 454003, China (e-mail: lixingwangbupt@gmail.com).
	
	Symeon Chatzinotas is with the Interdisciplinary Centre for Security, Reliability and Trust (SnT), University of Luxembourg, 1855 Luxembourg City, Luxembourg (e-mail: symeon.chatzinotas@uni.lu).
	
	(Corresponding author: Zhu Shoujin)
}

\vspace{-0.6cm}}%

\markboth{}
{ \MakeLowercase{\textit{}}} 
\maketitle

\begin{abstract}
This study investigates the synergy between rate-splitting multiple access (RSMA) and simultaneous transmitting and reflecting reconfigurable intelligent surfaces (STAR-RIS) as a unified framework to realize ubiquitous, intelligent, and resilient connectivity in future sixth-generation networks, while enhancing both spectral and energy efficiency. Specifically, in the STAR-RIS-assisted multi-user RSMA network under consideration, we develop an intelligent optimization strategy that jointly designs the active beamforming at the transmitter, the allocated transmission rate for the common stream, and the passive beamforming vectors for both transmission and reflection regions of the STAR-RIS, while accounting for transceiver hardware impairments and imperfect channel state information (CSI). In addition, system robustness is ensured by incorporating a bounded channel estimation error model that rigorously reflects CSI imperfections and ensures resilience against worst-case estimation errors. To tackle the highly non-convex problem, we propose an intelligent optimization algorithm that decouples the original problem into two sub-problems, which are then solved iteratively. Firstly, the active beamforming for both the common and private signals are obtained by reformulating the original non-convex problem into a tractable convex semi-definite programming (SDP) framework, leveraging successive convex approximation (SCA) and semi-definite relaxation (SDR) for enhanced computational efficiency. Secondly, the passive beamforming vectors for the transmission and reflection regions of the STAR-RIS are optimized through a convex SDP reformulation by exploiting SCA and SDR techniques. Additionally, when the resulting active or passive beamforming solutions are of higher rank, Gaussian randomization is employed to construct rank-one solutions. Finally, the effectiveness of the proposed optimization strategy is demonstrated through numerical simulations, which reveal significant performance gains over benchmark schemes and confirm rapid convergence.
	
\end{abstract}

\begin{IEEEkeywords} Rate-splitting multiple access (RSMA), reconfigurable intelligent surface (RIS), simultaneous transmitting and reflecting RIS (STAR-RIS), hardware impairments, channel estimation errors. 
\end{IEEEkeywords}

\IEEEpeerreviewmaketitle

\section{Introduction}

\IEEEPARstart{I} {n} recent years, the rollout of fifth-generation (5G) wireless networks has enabled substantial advancements in supporting diverse emerging applications, including virtual-reality, augmented-reality, remote healthcare, and autonomous transportation \cite{wang2022gcwcn, zhang20196g}. In spite of these developments, the rapid proliferation of internet-of-things (IoT) devices has significantly escalated data traffic, thereby revealing intrinsic limitations of 5G systems. These constraints are particularly evident in terms of  ubiquitous coverage, massive connectivity, spectral and energy efficiency, and operational robustness in heterogeneous environments \cite{saad2019vision}. In response to these emerging challenges, significant efforts are being made by both academia and industry to conceptualize and develop sixth-generation (6G) wireless networks. The primary goal is to enable ubiquitous, intelligent, and highly resilient connectivity capable of meeting the stringent performance demands of emerging next-generation applications and services, while simultaneously enhancing spectral and energy efficiency \cite{chen2020vision, wang2023road}. In this context, realizing the ambitious vision of 6G requires the integration of disruptive paradigms like rate-splitting multiple access (RSMA) \cite{mao2022rate, clerckx2023primer} and reconfigurable intelligent surface (RIS) \cite{di2020smart, liu2021reconfigurable}, both of which promise substantial gains in spectral-efficiency, energy-efficiency, coverage, and network connectivity.

RSMA has been identified as a promising physical-layer solution for non-orthogonal transmission, offering superior interference management compared to conventional multiple access schemes \cite{clerckx2023primer,clerckx2019rate,mao2019rate}. The fundamental concept of RSMA lies in its hybrid interference treatment, where part of the interference is decoded while the remaining portion is regarded as noise. To achieve this, the transmitter separates each user’s message into two components: a private segment, encoded independently for each user with a dedicated codebook, and a common segment, which combines information destined for several users and encodes it collectively into a unified stream using a commonly accessible codebook. At the receiver side, each user first retrieves the common data stream and applies successive interference cancellation (SIC) to eliminate its contribution from the received signal, and subsequently decodes its designated private stream while interpreting other users’ private signals as residual interference \cite{mao2022rate, clerckx2023primer}. Thus, by tuning the ratio of data splitting and the power allocation between common and private streams, RSMA is capable of operating along a spectrum ranging from treating interference purely as noise to fully decoding it with SIC \cite{camana2022rate}. Due to its flexible structure, RSMA bridges the gap between space-division multiple access (SDMA), which treats interference as noise, and non-orthogonal multiple access (NOMA), which attempts to fully decode all interfering signals. Thus, RSMA emerges as a balanced and adaptable strategy for next-generation communication networks \cite{clerckx2019rate,camana2022rate}. As a result, RSMA leverages its balanced transmission strategy to enhance spectral and energy efficiency \cite{mao2019rate,camana2022rate}, while ensuring improved robustness in multi-user networks operating under adverse conditions.

While RSMA offers a highly flexible and effective means of improving spectral and energy efficiency for future 6G networks, its performance can be significantly challenged by the propagation characteristics of high-frequency signals, which are prone to severe fading and limited penetration capability. In RSMA-enabled networks, system's performance deteriorates further if the direct communication link is obstructed by large environmental blockages, leading to coverage holes and potential information outages in dead zones. To mitigate these limitations and fully exploit the benefits of RSMA in next-generation wireless networks, the integration of RIS has emerged as a promising solution, offering the ability to reconfigure the wireless environment and enhance link reliability \cite{liu2021reconfigurable,wu2019towards}. Specifically, RIS technology can intelligently manipulate the wireless propagation environment, thus improving energy efficiency, coverage, and overall network connectivity \cite{asif2024securing}. In general, an RIS comprises a planar array containing numerous passive reflective elements. By controlling the reflection units through an integrated micro-controller, RIS can steer incoming signals toward desired directions, actively shaping the channel environment and mitigating coverage blind spots \cite{ihsan2022energy}. Owing to its passive nature, which requires minimal energy consumption beyond the micro-controller, RIS is not only cost-effective and easy to deploy but also demands very little maintenance.

Motivated by the aforementioned advantages of RIS, numerous studies have explored its integration with RSMA to further enhance network performance \cite{khisa2023energy,zhang2023energy,sun2023joint,soleymani2025rate,xu2024robust,pala2023spectral}. Specifically, authors in \cite{khisa2023energy}, developed a joint optimization framework to minimize the network energy-consumption by jointly designing the BS precoding vectors, determining the common stream split, allocating transmit power at the relay, scheduling time slots, and configuring RIS-based passive. In \cite{zhang2023energy}, the authors developed an optimization framework for an RIS-assisted RSMA system integrated with the SWIPT protocol to maximize energy efficiency by jointly designing the beamforming vectors, power-splitting (PS) ratios, common message rates, and discrete phase shifts. Further, the authors in \cite{sun2023joint} investigated an uplink RIS-assisted RSMA system and formulated an optimization problem to maximize sum-rate. The proposed design jointly optimizes transmit power allocation and passive beamforming, while satisfying the users’ quality-of-service (QoS) requirements \cite{sun2023joint}. Authors in \cite{soleymani2025rate} developed RSMA schemes for multiple-input multiple-output (MIMO) RIS-assisted broadcast channels under finite block-length coding. Their results revealed that combining RSMA with RIS yields substantial improvements in both spectral and energy efficiency. Moreover, the performance gains obtained from employing RSMA and RIS increased noticeably as the reliability and latency constraints became more stringent. In \cite{xu2024robust}, an optimization strategy was proposed to minimize the total transmit power in an RIS-aided symbiotic radio RSMA system. In particular, the total transmit power of the considered system was minimized by jointly optimizing the active-beamforming at the transmitter and the passive-beamforming at the receiver, while adhering to QoS and power budget requirements. Additionally, an alternating optimization algorithm was introduced for a multi-RIS-assisted RSMA system, in which the sum-rate was maximized by jointly optimizing the transmit beamforming at the base station (BS), passive beamforming at the RIS, and the block-length allocation of private and common data streams \cite{pala2023spectral}.

Nevertheless, the aforementioned studies \cite{khisa2023energy,zhang2023energy,sun2023joint,soleymani2025rate,xu2024robust,pala2023spectral} are limited to the conventional RIS architecture, which provides only half-space coverage. In particular, a traditional RIS is capable of either reflecting or refracting signals, but cannot simultaneously serve users located on both sides of the surface. To address this limitation, an advanced RIS architecture, referred to as simultaneous transmitting and reflecting RIS (STAR-RIS) \cite{liu2022star,mu2021simultaneously, zhang2022intelligent}, has been introduced. Unlike the conventional design, STAR-RIS enables full-space coverage by simultaneously reflecting and transmitting the incident signals, thereby supporting users on both sides of the surface. To this end, several studies have explored STAR-RIS-assisted RSMA frameworks to achieve improved coverage, capacity, and robustness in wireless networks. For instance, in \cite{katwe2023improved}, an alternating optimization framework was considered to maximize the sum-rate of a STAR-RIS aided RSMA uplink communication system by jointly optimizing the decoding order, power allocation, user fairness, and passive beamforming at the STAR-RIS. Accordingly, an alternating optimization strategy was proposed for an RSMA-enabled STAR-RIS assisted bistatic backscatter communication system, where the sum-rate was maximized by jointly optimizing the passive beamforming at the STAR-RIS and the power allocation at the backscatter tag under QoS and energy-harvesting constraints \cite{xiao2024joint}. Additionally, the authors in \cite{wang2025maximizing,wang2024secure} proposed secure transmission schemes for STAR-RIS assisted RSMA networks, in which the system’s sum secrecy rate was maximized through the joint optimization of active and passive beamforming. Further, authors in \cite{liu2025star} investigated an optimization framework for a STAR-RIS-assisted RSMA-enabled integrated sensing and communication system. In particular, the authors maximized the signal-to-interference-plus-noise ratio (SINR) by jointly optimizing the common rate allocation for RSMA users, as well as the active and passive beamforming vectors. Further, the study in \cite{meng2023sum} introduced a deep reinforcement learning model utilizing the proximal policy optimization algorithm to improve the sum-rate of a STAR-RIS-empowered RSMA communication network. Finally, the work in \cite{chang2024star} introduced an alternating optimization strategy to maximize the covert communication rate of an RSMA-enabled STAR-RIS-assisted covert communication system by jointly optimizing the common rate allocation, transmit precoding vectors, and the transmission and reflection coefficients of the STAR-RIS.

Although the aforementioned studies \cite{katwe2023improved,xiao2024joint,wang2025maximizing,wang2024secure,liu2025star,meng2023sum,chang2024star} demonstrated the potential benefits of STAR-RIS-assisted RSMA systems through efficient resource management strategies, they typically assumed perfect transceiver hardware design. In practical wireless communication networks, hardware performance gradually degrades under environmental factors, causing signal distortions that deviate from the intended design. Such degradation affects key components, including power amplifiers, digital-to-analog converters, and oscillators, ultimately giving rise to hardware impairments (HIs)\cite{asif2024securing,li2019residual,asif2024leveraging}. In addition, analytical and experimental studies have shown that the distortion noise resulting from the HIs can be accurately modeled as additive Gaussian noise, with its variance being proportional to the signal power. Thus, achieving reliable performance in practical systems requires explicitly accounting for HIs in the signal model and incorporating appropriate compensation techniques within the optimization framework. In this context, the study in \cite{wang2025robust} considered an alternating optimization algorithm for a STAR-RIS-assisted RSMA system under HIs. However, their work is subject to the following limitations: 1) Firstly, their work focused solely on maximizing the system sum-rate, neglecting users’ QoS constraints during both active and passive beamforming optimization for the sake of simplicity. However, overlooking QoS requirements may lead to solutions that fail to meet essential system requirements; 2) Secondly, they assumed the availability of perfect channel state information (CSI) for all links, which is often impractical in complex propagation environments. Consequently, their proposed optimization framework suffers from limited robustness.

Motivated by these limitations, we design a robust resource allocation scheme for STAR-RIS-enabled RSMA network under HIs, which explicitly incorporates channel estimation errors while ensuring QoS satisfaction, power budget adherence, common rate design, and energy conservation constraints of the system. Consequently, the major contributions of this work are organized as follows:
\begin{itemize}
	\item We consider a multi-user downlink communication system assisted by a STAR-RIS and operating under the RSMA protocol. In this scenario, the direct transmission link between the BS and the users is assumed to be blocked by substantial environmental obstructions, such that the users located in both the transmission and reflection regions are served exclusively through the STAR-RIS. Furthermore, to capture more practical system characteristics, it is considered that both the transmitter and receiver are subject to HIs arising from non-ideal hardware design.
	
	\item Further, to guarantee system robustness, the sum-rate maximization problem is formulated under imperfect CSI at the BS, and a bounded channel uncertainty model is adopted to characterize the worst-case effect of CSI errors.
	
	\item Next, a unified optimization strategy is developed to enhance the system sum-rate through the simultaneous design of active beamforming, adjustment of the actual transmission rate of the common stream, and configuration of the STAR-RIS transmission and reflection coefficients, subject to QoS guarantees, transmit power limits, common rate requirements, and energy conservation constraints.
	
	\item Subsequently, to tackle the non-convex nature of the formulated problem, we design an intelligent alternating optimization framework. In the first stage, the active beamforming vectors are optimized by reformulating the problem into a semi-definite programming (SDP) problem by leveraging successive convex approximation (SCA) and semi-definite relaxation (SDR) techniques. In the second stage, the passive beamforming vectors corresponding to the transmission and reflection regions of the STAR-RIS are optimized through a convex SDP reformulation by exploiting SCA and SDR techniques. Moreover, when the obtained active or passive beamforming solutions are of higher rank, the Gaussian randomization method is applied to recover rank-one solutions.
	
	\item Finally, the effectiveness of the proposed optimization strategy is validated through extensive numerical simulations, which demonstrate considerable performance gains over benchmark schemes and exhibit fast convergence.
	
\end{itemize}

 \begin{figure}[!t]
	\centering
	\includegraphics [width=0.46\textwidth]{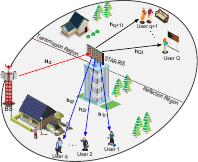}
	\caption{System model.}
	\label{f1}
\end{figure}

\section{System Model and Optimization Problem Formulation}
In this work, we investigate an RSMA-based multi-user downlink system where a BS realized with $N$ antennas, configured as a uniform linear array (ULA), serves $Q$ single-antenna users in the presence of channel estimation errors and transceiver imperfections. As depicted in Fig.~\ref{f1}, it is assumed that the direct communication link between the BS and the users is blocked by environmental obstacles, and a STAR-RIS with $M$ elements, configured in a uniform rectangular array (URA) is deployed to assist the users by simultaneously operating in both the transmission and reflection regions \cite{asif2024securing}.  

Next, the channel gain from the BS to the STAR-RIS is represented by $\mathbf{H}_{d} \in \mathbb{C}^{M \times N}$, while the channel gain from the STAR-RIS to the $q$-th user is denoted by $\mathbf{h}_{qx} \in \mathbb{C}^{M \times 1}$. Here, $\forall x \in \{r,t\}$ indicates whether the RSMA user is located in the reflection region ($x=r$) or the transmission region ($x=t$). Additionally, all the wireless channels, i.e., $\mathbf{H}_{d} \in \mathbb{C}^{M \times N}$ and $\mathbf{h}_{qx} \in \mathbb{C}^{M \times 1}$ for each user $q \in \mathcal{Q} = \{1,2,\ldots,Q\}$, are assumed to follow the Rician fading distribution. Accordingly, the Rician fading model for any channel $\tilde{\psi} \in \{{\mathbf{H}_{d}, \mathbf{h}_{qx}}\}$ is expressed as
\begin{equation}
	\tilde{\psi} = \sqrt{\tilde{\vartheta}_{i}} \left( \sqrt{\frac{\alpha_{i}}{1+\alpha_{i}}} \tilde{\psi}_{\mathrm{LOS}} + \sqrt{\frac{1}{1+\alpha_{i}}} \tilde{\psi}_{\mathrm{NLOS}} \right),
	\label{1}
\end{equation}
where $\alpha_{i}$ and $\tilde{\vartheta}_{i}$, $\forall i \in \{{d, qx}\}$, denote the Rician factor and the large-scale path loss corresponding to the channel $\tilde{\psi}$, respectively. Moreover, $\tilde{\psi}_{\mathrm{LOS}}$ and $\tilde{\psi}_{\mathrm{NLOS}}$ represent the line-of-sight (LOS) and non-line-of-sight (NLOS) components of $\tilde{\psi}$, respectively.

Further, by leveraging the RSMA protocol, each user’s message $m_q$ is decomposed into two parts: a common message $m_q^0$ and a private message $m_q^p$. The common messages of all users are jointly encoded to generate a single common stream $\hat{s}_0$, whereas the private messages are individually encoded into $Q$ distinct private streams, denoted by ${\hat{s}_1, \hat{s}_2, \ldots, \hat{s}_Q}$, with $q \in \mathcal{Q} = \{1,2,\ldots,Q\}$. Hence, the transmitted signal is given by
\begin{equation}
	\mathbf{x}=\sum_{q=1}^Q \mathbf{w}_q \hat{s}_q+\mathbf{w}_c \hat{s}_c+\boldsymbol{\Xi}_t, \label{2}
\end{equation}
where $\mathbf{w}_q$ $\mathbb{C}^{N \times 1}$ and $\mathbf{w}_c$ $\mathbb{C}^{N \times 1}$ denote transmit precoding vectors for private and common signals, respectively. Let $\mathbb{E}[|\hat{s}_{0}|^2]=1$ and $\mathbb{E}[|\hat{s}_{q}|^2]=1$, $\forall q \in \mathcal Q$. Further, the HIs at the transmitter can be modeled by the random variable $\boldsymbol{\Xi}_t \sim \mathcal{CN}\ \!\Big(\mathbf{0}, \, \kappa_t \, \widetilde{\mathrm{diag}} \ \!\Big(\sum_{q=1}^{Q} \mathbf{w}_{q}\mathbf{w}_{q}^{H} + \mathbf{w}_{c}\mathbf{w}_{c}^{H}\Big)\Big)$ \cite{bjornson2014massive,shen2020beamforming,zhang2023robust}, which follows a circularly symmetric complex Gaussian distribution with zero mean. The variance of $\boldsymbol{\Xi}_t$ is proportional to the transmit power across antennas, where the parameter $\kappa_t \in (0,1)$ specifies the ratio of transmit distortion noise power to the transmit signal power \cite{shen2020beamforming,zhang2023robust}. 

Next, the passive beamforming vector of the STAR-RIS, $\hat{\mathbf{e}}_{x}$, with $x \in \{t,r\}$ representing the transmission and reflection regions, is given by $\hat{\mathbf{e}}_{x} = \big[\sqrt{\zeta_1^{x}} e^{j \vartheta_1^{x}}, \sqrt{\zeta_2^{x}} e^{j \vartheta_2^{x}}, \ldots, \sqrt{\zeta_M^{x}} e^{j \vartheta_M^{x}}\big]^T$, $\forall m\in \mathcal M = \{1,2,..., M\}$. Herein, $\sqrt{\zeta_m^{x}} \in [0,1]$ and $\vartheta_m^{x} \in [0,2\pi]$ denote the amplitude and phase-shift of the $m$-th STAR-RIS element, for all $m \in \mathcal{M} = \{1,2,\ldots,M\}$ \cite{asif2024securing}.

Accordingly, the received signal at the $q$-th user, located either in the transmission or reflection region of the STAR-RIS, can be written as
 \begin{align}
 	& y_{q,x}= \hat{y}_{q,x} + \Xi_{r,q}, \forall q\in \mathcal Q, {x}\in \{{t},{r}\},\label{3}
 \end{align}
where $\Xi_{r,q} \sim \mathcal{CN}\big(0, \kappa_r \mathbb{E}[|\hat{y}_{q,x}|^2]\big)$ \cite{shen2020beamforming} denotes the the HIs at $q$-th receiver, positioned within the transmission or reflection region of the STAR-RIS. $\kappa_r \in (0,1)$ \cite{zhang2023robust} denotes the ratio of distorted noise power to undistorted signal power at $q$-th user. 

Next, we define the equivalent channel gain for the $q$-th user in the transmission or reflection region of the STAR-RIS as $\hat{\mathbf{h}}_{qx} = \mathbf{H}_{d}^H \mathbf \Theta_x \mathbf h_{qx}$, where $\mathbf{\Theta}_x = \operatorname{diag}(\hat{\mathbf{e}}_{x})$ represents the scattering matrix of the STAR-RIS. Then, the undistorted received signal component $\hat{y}_{q,x}$ is given by
 \begin{align}
	& \hat{y}_{q,x}= \hat{\mathbf{h}}^H_{qx}\mathbf x + {n}_q, \label{4}
\end{align} 
where ${n}_q \sim (0, \sigma_{q}^2)$ denotes the additive white Gaussian noise at $q$-th receiver. Thus, the achievable rate at $q$-th user for decoding the common message can be expressed as
\begin{figure*} 
	\begin{align} \label{5} 
		\scalemath{0.90}{\operatorname{\Gamma}^c_{q,x}=\frac{|  \hat{\mathbf{h}}^H_{qx} \mathbf{w}_c|^2}{\sum_{j=1}^Q| \hat{\mathbf h}_{qx}^H \mathbf{w}_j |^2+\kappa_r\left(\sum_{j=1}^Q|\hat{\mathbf h}^H_{qx} \mathbf{w}_j |^2+| \hat{\mathbf{h}}^H_{qx}  \mathbf{w}_c|^2\right)+\left(1+\kappa_r\right) \kappa_t \hat{\mathbf{h}}_{qx}^H \widetilde{\operatorname{diag}}\left(\sum_{j=1}^Q \mathbf{w}_j \mathbf{w}_j^H+\mathbf{w}_c \mathbf{w}_c^H\right) \hat{\mathbf{h}}_{qx}+\left(1+\kappa_r\right) \sigma_q^2}},
	\end{align} 
\end{figure*}
\begin{figure*}
	\begin{align} \label{6}
	\scalemath{0.90}{\operatorname{\Gamma}^p_{q,x}=\frac{| \hat{\mathbf{h}}^H_{qx} \mathbf{w}_q|^2}{\sum_{i \neq q}| \hat{\mathbf{h}}^H_{qx}\mathbf{w}_i|^2+\kappa_r\left(\sum_{j=1}^Q| \hat{\mathbf h}^H_{qx} \mathbf w_j|^2+| \hat{\mathbf{h}}^H_{qx} \mathbf{w}_c|^2\right)+\left(1+\kappa_r\right) \kappa_t \hat{\mathbf{h}}^H_{qx} \widetilde{\operatorname{diag}}\left(\sum_{j=1}^Q \mathbf{w}_j \mathbf{w}_j^H+\mathbf{w}_c \mathbf{w}_c^H\right) \hat{\mathbf{h}}_{qx}+\left(1+\kappa_r\right) \sigma_q^2}},
	\end{align}\hrulefill
\end{figure*}
\begin{equation}
	R^c_{q,x}=\log _2\left(1+\operatorname{\Gamma}^c_{q,x}\right), \label{7}
\end{equation}
where $\Gamma^c_{q,x}$ is given by Eq.~\eqref{5}. After decoding the common message, the receiver applies SIC to remove its contribution from the received signal. Thereafter, each RSMA user decodes its respective private stream, regarding the remaining users’ private signals as interference \cite{asif2024leveraging}. Consequently, the achievable rate of the private message is given as
\begin{equation}
	R^p_{q,x}=\log _2\left(1+\operatorname{\Gamma}^p_{q,x}\right), \label{8}
\end{equation}
where $\Gamma^p_{q,x}$ is given by Eq.~\eqref{6}.

Next, to account for channel estimation errors, the equivalent channel of the $q$-th user in region $x$ is modeled as $\overline{\mathbf{h}}_{qx} = \hat{\mathbf{h}}_{qx} + \Delta \hat{\mathbf{h}}_{qx}$, where $\hat{\mathbf{h}}_{qx}$ denotes the estimated CSI and $\Delta \hat{\mathbf{h}}_{qx}$ represents the estimation error following a complex Gaussian distribution \cite{zhang2023robust,li2022robust}. The error is constrained by $\|\Delta \hat{\mathbf{h}}_{qx}\| \leq \varepsilon_{qx}$ \cite{li2022robust}, and we assume that $\overline{\mathbf{h}}_{qx}$ and $\Delta \hat{\mathbf{h}}_{qx}$ are statistically independent. Accordingly, the corresponding channel covariance matrix can be expressed as $\overline{\mathbf{H}}_{qx} = \hat{\mathbf{H}}_{qx} + \Delta \hat{\mathbf{H}}_{qx}$, where $\overline{\mathbf{H}}_{qx} = \mathbb{E}\{\overline{\mathbf{h}}_{qx}\overline{\mathbf{h}}_{qx}^H\}$, $\hat{\mathbf{H}}_{qx} = \mathbb{E}\{\hat{\mathbf{h}}_{qx}\hat{\mathbf{h}}_{qx}^H\}$, and $\Delta \hat{\mathbf{H}}_{qx} = \mathbb{E}\{\Delta \hat{\mathbf{h}}_{qx}\Delta \hat{\mathbf{h}}_{qx}^H\}$ correspond to the covariance matrices of the true channel, the estimated channel, and the associated estimation error, respectively.

Thus, under channel estimation errors, the updated values of $\operatorname{\Gamma}^c_{q,x}$ and $\operatorname{\Gamma}^p_{q,x}$ are given by \eqref{9} and \eqref{10}, respectively, where $\tilde{\sigma}_q^2 = (1+\kappa_r)\sigma_q^2$. Accordingly, \eqref{7} and \eqref{8} can be updated as
\begin{figure*} 
	\begin{align} \label{9}
		\scalemath{0.90}{\overline{\Gamma}^{c}_{q,x} 
			= \frac{ \operatorname{tr}\!\Big(\mathbf w_c^H\big(\hat{\mathbf{H}}_{qx} + \Delta \hat{\mathbf{H}}_{qx}\big) \mathbf w_c\Big) }
			{\operatorname{tr}\!\Bigg(
				\Bigg(
				(1+\kappa_r)\sum_{j=1}^Q \mathbf{w}_j \mathbf{w}_j^{\!H}
				\;+\; \kappa_r\, \mathbf{w}_c \mathbf{w}_c^{\!H}
				\;+\; (1+\kappa_r)\kappa_t\, \widetilde  {\operatorname{diag}}\!\Big(\sum_{j=1}^Q \mathbf{w}_j \mathbf{w}_j^{\!H} + \mathbf{w}_c \mathbf{w}_c^{\!H}\Big)
				\Bigg) \Big(\hat{\mathbf{H}}_{qx} + \Delta \hat{\mathbf{H}}_{qx}\Big)
				\Bigg)
				+ \tilde{\sigma}_q^2 },}
	\end{align}
\end{figure*}
\begin{figure*} 
	\begin{align} \label{10}
		\scalemath{0.90}{\overline{\Gamma}^{p}_{q,x} 
			= \frac{ \operatorname{tr}\!\Big(\mathbf w_q^H\big(\hat{\mathbf{H}}_{qx} + \Delta \hat{\mathbf{H}}_{qx}\big) \mathbf w_q\Big) }
			{\operatorname{tr}\!\Bigg(
				\Bigg(
				\sum_{i \neq q} \mathbf{w}_i \mathbf{w}_i^{\!H}+\kappa_r\sum_{j=1}^Q \mathbf{w}_j \mathbf{w}_j^{\!H}
				\;+\; \kappa_r\, \mathbf{w}_c \mathbf{w}_c^{\!H}
				\;+\; (1+\kappa_r)\kappa_t\, \widetilde  {\operatorname{diag}}\!\Big(\sum_{j=1}^Q \mathbf{w}_j \mathbf{w}_j^{\!H} + \mathbf{w}_c \mathbf{w}_c^{\!H}\Big)
				\Bigg) \Big(\hat{\mathbf{H}}_{qx} + \Delta \hat{\mathbf{H}}_{qx}\Big)
				\Bigg)
				+ \tilde{\sigma}_q^2 },}
	\end{align}\hrulefill
\end{figure*}
\begin{equation}
	\overline{R}^c_{q,x}=\log _2\left(1+\operatorname{\overline{\Gamma}}^c_{q,x}\right), \label{11}
\end{equation}
\begin{equation}
	\overline{R}^p_{q,x}=\log _2\left(1+\operatorname{\overline{\Gamma}}^p_{q,x}\right). \label{12}
\end{equation}

Next, for the considered system, the common data stream can only be decoded if the following condition is satisfied
\begin{align}
	\sum\limits_{{\substack{q = 1}}}^Q {r}_q^c	\leq \log _2\left(1+\operatorname{\overline{\Gamma}}^c_{q,x}\right), \label{13}
\end{align}
where ${r}_q^c$ denotes the fraction of the common rate assigned to the $q$-th user. Accordingly, the total achievable system sum-rate is expressed as
	\begin{align}
	\overline{R}= & \sum\limits_{{\substack{q = 1}}}^Q \Big({r}^c_q + \overline{R}^p_{q,x}\Big), \forall q\in \mathcal Q.  \label{14}
\end{align}

Consequently, the optimization problem for the considered system is formulated as
\begin{subequations}\label{P1}
	\begin{align}
		\text{\textbf{(P1)}}	& \mathop {\max }\limits_{({r}^c_q, \mathbf {w}_{c}, \mathbf {w}_{q}, \mathbf \Theta_x)} \    \sum\limits_{{\substack{q=1}}}^Q \Big({r}^c_q + \overline{R}^p_{q,x}\Big), \label{15a}  \\
		s.t.\ C_1:\ & {r}^c_q +\overline{R}^p_{q,x}  \geq  {R}_{min}, \forall q\in \mathcal Q, \label{15b} \\
		\ C_2:\ &  \sum\limits_{{\substack{q=1}}}^Q  {r}^c_q  \leq \overline{R}^c_{q,x}, \forall q\in \mathcal Q, \label{15c}\\
		\ C_3:\ & \left\|\mathbf w_c\right\|^2+\sum\limits_{{\substack{q=1}}}^Q \left\|\mathbf w_q\right\|^2 \leq P_{budg}, \label{15d}\\
		\ C_4:\ & \zeta^{t}_{m}+ \zeta^{r}_{m} = 1, \forall m\in \mathcal M, \label{15e} \\ 
		\ C_5: \ &\vartheta^x_m \in [0, 2\pi], \forall m\in \mathcal M, \label{15f} 
	\end{align}
\end{subequations}
where $R_{\min}$ represents the minimum rate required to satisfy the QoS constraint, and $P_{\text{budg}}$ denotes the maximum available power budget. Constraints $C_1$ and $C_2$ ensure the QoS requirements and the successful decoding of the common message, respectively. Constraint $C_3$ regulates the transmission power according to the total system power budget. Finally, $C_4$ and $C_5$ represent the energy conservation and phase-shift constraints of the STAR-RIS, respectively.
\section{Proposed Solution for the Optimization Problem}
Given the inherent complexity of the optimization problem \textbf{(P1)}, we begin by reformulating it into a more tractable equivalent form. In this regard, we present \textbf{Theorem 1} as follows:\\
\textbf{Theorem 1:} 
Let $\mathbf{\widetilde{A}}$ and $\mathbf{\widetilde{B}}$ be Hermitian matrices, where $\mathbf{\widetilde{A}}$ is a positive semi-definite (PSD) matrix with rank equal to one. If $\mathbf{\widetilde{B}}$ is subject to the constraint $\|\mathbf{\widetilde{B}}\| \leq \varepsilon^2$, 
then the following equality holds
\begin{equation}
	\max_{\|\mathbf{\widetilde{B}}\| \leq \varepsilon^2} \operatorname{tr}(\mathbf{\widetilde{A}}\mathbf{\widetilde{B}})
	= \varepsilon^2 \operatorname{tr}(\mathbf{\widetilde{A}}). \label{16}
\end{equation}

\noindent \textbf{Proof:} Assume that $\mathbf{\widetilde{S}}$ and $\mathbf{\widetilde{U}}$ are complex matrices such that $\|\mathbf{\widetilde{S}}\| \leq 1$. In this case, the maximum achievable value of the inner product $\langle \mathbf{\widetilde{S}}, \mathbf{\widetilde{U}} \rangle$ is attained at the dual norm of $\mathbf{\widetilde{U}}$, that is
\begin{equation}
	\max_{\|\mathbf{\widetilde{S}}\| \leq 1} \langle \mathbf{\widetilde{S}}, \mathbf{\widetilde{U}} \rangle 
	= \|\mathbf{\widetilde{U}}\|_{dual}. \label{17}
\end{equation}

Then, based on \eqref{17}, the following inequality is obtained:
\begin{equation}
	\operatorname{tr}(\mathbf{\widetilde{U}}^H \mathbf{\widetilde{S}}) \leq 
	\|\mathbf{\widetilde{S}}\| \, \|\mathbf{\widetilde{U}}\|_{\text{dual}}, \label{18}
\end{equation}
where $\|\cdot\|_{\text{dual.}}$ denotes the corresponding dual norm. 
Accordingly, for Hermitian matrices  $\mathbf{\widetilde{A}}$ and $\mathbf{\widetilde{B}}$ with $\|\mathbf{\widetilde{B}}\| \leq \varepsilon^2$, we can further write
\begin{equation}
	\operatorname{tr}(\mathbf{\widetilde{A}} \mathbf{\widetilde{B}}) \leq \|\mathbf{\widetilde{B}}\| \|\mathbf{\widetilde{A}}\|_{dual} \leq \varepsilon^2 \|\mathbf{\widetilde{A}}\|_{dual}. \label{19}
\end{equation} 
Thus,
\begin{equation}
	\max _{\|\mathbf{\widetilde{B}}\| \leq \varepsilon^2} \operatorname{tr}(\mathbf{\widetilde{A}} \mathbf{\widetilde{B}})=\varepsilon^2 \|\mathbf{\widetilde{A}}\|_{dual}. \label{20}
\end{equation} 

By evaluating the spectral norm of $\mathbf{\widetilde{B}}$, it follows that $\xi_{\max}(\mathbf{\widetilde{B}}) \leq \varepsilon^2$, where $\xi_{\max}$ represents the maximum eigenvalue of $\mathbf{\widetilde{B}}$. Given that the nuclear norm is the dual of the spectral norm, the dual norm associated with $\mathbf{\widetilde{A}}$ can therefore be written as
\begin{equation}
	\|\mathbf{\widetilde{A}}\|_{\text{dual}} = \|\mathbf{\widetilde{A}}\|_{\text{nuclear}} 
	= \sum_{i} \xi_i. \label{21}
\end{equation} 

Given that $\mathbf{\widetilde{A}}$ is a PSD matrix with rank-one, its nuclear norm reduces to its trace, i.e.,	$\|\mathbf{\widetilde{A}}\|_{\text{dual}} = \operatorname{tr}(\mathbf{\widetilde{A}})$. Accordingly, Eq.~\eqref{20} can be reformulated as
\begin{equation}
	\max _{\|\mathbf{\widetilde{B}}\| \leq \varepsilon^2} \operatorname{tr}(\mathbf{\widetilde{A}}\mathbf{\widetilde{B}})=\varepsilon^2 \operatorname{tr}(\mathbf{\widetilde{A}}). \label{22}
\end{equation} 
Hence, the proof is complete.

Then, by adopting \textbf{Theorem 1}, we can write as follows
\begin{equation}
	\max _{\|\Delta \hat{\mathbf{H}}_{qx}\| \leq \varepsilon_{qx}^2} \operatorname{tr}(\Delta \hat{\mathbf{H}}_{qx} \mathbf W_c)=\varepsilon_{qx}^2 \operatorname{tr}(\mathbf{W_c}) \label{23}
\end{equation} 
and,
\begin{equation}
	\max _{\|\Delta \hat{\mathbf{H}}_{qx}\| \leq \varepsilon_{qx}^2} \operatorname{tr}(\Delta \hat{\mathbf{H}}_{qx} \mathbf W_q)=\varepsilon_{qx}^2 \operatorname{tr}(\mathbf{W_q}). \label{24}
\end{equation} 

Furthermore, we consider the system under a worst-case assumption, where the desired signal channels are overestimated, leading to $\Delta \hat{\mathbf{H}}_{qx} = -\varepsilon_{qx}^2 \mathbf{I}$, while the interference channels are underestimated, yielding $\Delta \hat{\mathbf{H}}_{qx} = \varepsilon_{qx}^2 \mathbf{I}$ \cite{li2022robust, zheng2023zero}. 

Next, we define $\mathbf{W}_c = \mathbf{w}_c \mathbf{w}_c^H$ and 
$\mathbf{W}_q = \mathbf{w}_q \mathbf{w}_q^H$ as two PSD matrices, each of rank one, i.e., $\operatorname{rank}(\mathbf{W}_c) = 1$ and $\operatorname{rank}(\mathbf{W}_q) = 1$. Accordingly, \eqref{11} and \eqref{12} can be updated as
\begin{equation}
	\widetilde{R}^c_{q,x}=\log _2\left(1+\frac{ \operatorname{tr}\Bigg(\Big(\hat{\mathbf{H}}_{qx}-\varepsilon_{qx}^2 \mathbf{I}\Big)\mathbf{W}_c \Bigg) }
	{\operatorname{tr}\!\Bigg(
		 \Big(\hat{\mathbf{H}}_{qx}+\varepsilon_{qx}^2 \mathbf{I}\Big) \Big(\mathbf \Pi_1+ \mathbf \Pi_2\Big)
		\Bigg)
		+ \tilde{\sigma}_q^2 }\right), \label{25}
\end{equation}
\begin{equation}
	\widetilde{R}^p_{q,x}=\log _2\left(1+\frac{ \operatorname{tr}\Bigg(\Big(\hat{\mathbf{H}}_{qx}-\varepsilon_{qx}^2 \mathbf{I}\Big)\mathbf{W}_q \Bigg) }
	{\operatorname{tr}\!\Bigg(
		\Big(\hat{\mathbf{H}}_{qx}+\varepsilon_{qx}^2 \mathbf{I}\Big) \Big(\mathbf \Pi_2+ \mathbf \Pi_3\Big)
		\Bigg)
		+ \tilde{\sigma}_q^2 }\right), \label{26}
\end{equation}
where $\mathbf \Pi_1=
(1+\kappa_r)\sum_{j=1}^Q \mathbf W_j + \kappa_r \mathbf W_c$, $\mathbf \Pi_2= (1+\kappa_r)\kappa_t\, \widetilde  {\operatorname{diag}}\! \ \Big(\sum_{j=1}^Q \mathbf W_j + \mathbf W_c\Big)$, and $\mathbf \Pi_3 = 				\sum_{i \neq q} \mathbf{W}_i + \kappa_r\sum_{j=1}^Q \mathbf{W}_j+ \kappa_r \mathbf{W}_c$.

Consequently, the optimization problem \textbf{(P1)} can be reformulated as follows
\begin{subequations}\label{P2}
	\begin{align}
		\text{\textbf{(P2)}}	& \mathop {\max }\limits_{( r^c_q, \mathbf {W}_{c}, \mathbf {W}_{q}, \mathbf \Theta_x)} \    \sum\limits_{{\substack{q=1}}}^Q \Big(r^c_q + \widetilde{R}^p_{q,x}\Big), \label{27a}  \\
		s.t.\ \ & r^c_q + \widetilde{R}^p_{q,x}  \geq {R}_{min}, \forall q\in \mathcal Q, \label{27b} \\
		\ \ & \sum\limits_{{\substack{q=1}}}^Q  r^c_q  \leq \widetilde{R}^c_{q,x}, \forall q\in \mathcal Q, \label{27c}\\
		\ & \operatorname{tr}(\mathbf W_c)+\sum\limits_{{\substack{q=1}}}^Q \operatorname{tr}(\mathbf W_q) \leq P_{budg.}, \label{27d}\\
		\ \ & \mathbf W_c \succeq 0, \ \mathbf W_q \succeq 0, \ \forall q\in \mathcal Q, \label{27e}\\ 
		\ \ & \operatorname{rank}(\mathbf W_c)=1, \ \operatorname{rank}(\mathbf W_q)=1,  \forall q\in \mathcal Q. \label{27f}\\
		\  \ & \eqref{15e}, \eqref{15f}. \label{27g} 
	\end{align}
\end{subequations}

It should be noted that the coupling among optimization variables, $r^c_q$, $\mathbf{W}{c}$, $\mathbf{W}{q}$, and $\boldsymbol{\Theta}_x$, renders problem \textbf{(P2)} highly intricate and inherently non-convex. To address the non-convex nature of the problem, we utilize an alternating optimization strategy, wherein the original problem is decomposed into several sub-problems that are iteratively solved.

\subsection{Optimization of $\mathbf{w}_{c}$ and  $\mathbf{w}_{q}$: Stage 1}
Given that the passive beamforming matrix $\mathbf{\Theta}_x$ of the STAR-RIS is fixed, the optimization problem for determining the transmit precoding vectors $\mathbf{w}_{c}$ and $\mathbf{w}_{q}$ can be formulated as
\begin{subequations}\label{P3}
	\begin{align}
		\text{\textbf{(P3)}}	& \mathop {\max }\limits_{( r^c_q, \mathbf {W}_{c}, \mathbf {W}_{q})} \    \sum\limits_{{\substack{q=1}}}^Q \Big(r^c_q + \widetilde{R}^p_{q,x}\Big), \label{28a}  \\
		s.t.\ \ & \eqref{27b} - \eqref{27f} \label{28b} 
	\end{align}
\end{subequations}

Since the problem $\textbf{(P3)}$ is inherently non-convex, as its non-convexity originates from both the objective function and the associated constraints. In order to trace the convexity of the objective function expressed in \eqref{28a}, we introduce a slack variable $\Psi^p_{q,x}$, which is defined as follows
\begin{subequations}
	\begin{align}
		& r^c_q + \Psi^p_{q,x} \geq {R}_{min},\label{29a} 
	\end{align}
\begin{align}
	&\log_{2}\Biggl( \operatorname{tr}\Big(\Big(\hat{\mathbf{H}}_{qx}-\varepsilon_{qx}^2 \mathbf{I}\Big)\mathbf{W}_q \Big)+ \operatorname{tr}\!\Big(
	\Big(\hat{\mathbf{H}}_{qx}+\varepsilon_{qx}^2 \mathbf{I}\Big)\nonumber\\
	& \Big(\mathbf \Pi_2+ \mathbf \Pi_3\Big)
	\Big)
	+ \tilde{\sigma}_q^2  \Biggr) - \log_{2}\Biggl( \operatorname{tr}\!\Big(
	\Big(\hat{\mathbf{H}}_{qx}+\varepsilon_{qx}^2 \mathbf{I}\Big) \nonumber\\
	&\Big(\mathbf \Pi_2+ \mathbf \Pi_3\Big)
	\Big)
	+ \tilde{\sigma}_q^2 \Biggr) \geq \Psi^p_{q,x}, \forall q\in \mathcal Q  \label{29b}
\end{align}
\end{subequations}

Next, to facilitate the convexity analysis of \eqref{29b}, we introduce a slack variable vector $\boldsymbol{\beta}{q} = [\beta{q,1}, \beta_{q,2}]^T$ defined as
\begin{subequations}
		\begin{align}
		\operatorname{tr}\Bigg(\Big(\hat{\mathbf{H}}_{qx}&-\varepsilon_{qx}^2 \mathbf{I}\Big)\mathbf{W}_q \Bigg)+ \operatorname{tr}\!\Bigg(
		\Big(\hat{\mathbf{H}}_{qx}+\varepsilon_{qx}^2 \mathbf{I}\Big) \Big(\mathbf \Pi_2+ \mathbf \Pi_3\Big)
		\Bigg) \nonumber\\
		&
		+ \tilde{\sigma}_q^2  \geq \beta_{q,1}, \forall q\in \mathcal Q,\label{30a} 
	\end{align}
		\begin{align}
		& \operatorname{tr}\!\Bigg(
		\Big(\hat{\mathbf{H}}_{qx}+\varepsilon_{qx}^2 \mathbf{I}\Big)\Big(\mathbf \Pi_2+ \mathbf \Pi_3\Big)
		\Bigg)
		+ \tilde{\sigma}_q^2  \leq \beta_{q,2},\label{30b} 
	\end{align}
		\begin{align}
		&\log_{2}\Bigl( \beta_{q,1} \Bigr) - \log_{2}\Bigl(\beta_{q,2}\Bigr) \geq \Psi^p_{q,x}.  \label{30c}
	\end{align}
\end{subequations}

Since the left-hand side of \eqref{30c} remains non-convex owing to the presence of the second logarithmic term. Hence, to linearize the non-convex term, we utilize the first-order Taylor approximation as given below
\begin{align}
	&\log_{2}\Bigl( \beta_{q,1} \Bigr) - 	\Biggl(\log_{2}\big(\beta_{q,2}^{(r)}\big) +\frac{\beta_{q,2}-\beta_{q,2}^{(r)}}{\ln (2) \beta_{q,2}^{(r)}} \Biggr) \geq \Psi^p_{q,x},  \label{31}
\end{align}
where $\beta_{q,2}^{(r)}$ denotes the value of $\beta_{q,2}$ in $r$-th iteration. Further, to facilitate the convexity of \eqref{27c}, we introduce a slack variable vector $\boldsymbol{\upsilon}_q = [\upsilon_{q,1}, \upsilon_{q,2}]^{T}$, defined as follows
\begin{align}
	&\sum\limits_{{\substack{q = 1}}}^Q {r}_q^c \leq (\upsilon_{q,1} - \upsilon_{q,2}) ,  \label{32}
\end{align}
\begin{align}
	\log_{2}\Biggl(& \operatorname{tr}\Big(\Big(\hat{\mathbf{H}}_{qx}-\varepsilon_{qx}^2 \mathbf{I}\Big)\mathbf{W}_c \Big) + \operatorname{tr}\!\Big(
	\Big(\hat{\mathbf{H}}_{qx}+\nonumber \\
	&\varepsilon_{qx}^2 \mathbf{I}\Big) \Big(\mathbf \Pi_1+ \mathbf \Pi_2\Big)
	\Big)
	+ \tilde{\sigma}_q^2 \Biggr) \geq \upsilon_{q,1},  \label{33}
\end{align}
\begin{align}
	& \log_{2}\Biggl( \operatorname{tr}\!\Big(
	\Big(\hat{\mathbf{H}}_{qx}+\varepsilon_{qx}^2 \mathbf{I}\Big) \Big(\mathbf \Pi_1+ \mathbf \Pi_2\Big)
	\Big) + \tilde{\sigma}_q^2\Biggr) \leq \upsilon_{q,2}.  \label{34}
\end{align}

Next, to track the convexity of \eqref{34}, we introduce a slack variable $\kappa_q$ as follows
\begin{subequations}
	\begin{align}
	& \log_{2}\big(\kappa_q \big) \leq \upsilon_{q,2}, \label{35a}
	\end{align}
	\begin{align}
		&\operatorname{tr}\!\Bigg(
		\Big(\hat{\mathbf{H}}_{qx}+\varepsilon_{qx}^2 \mathbf{I}\Big) \Big(\mathbf \Pi_1+ \mathbf \Pi_2\Big)
		\Bigg) + \tilde{\sigma}_q^2 \leq \kappa_q.  \label{35b}
	\end{align}
\end{subequations}

Since \eqref{35a} is non-convex, we tackle it using the SCA approach given as
\begin{align}
	& \Biggl(\log_{2}\big(\kappa_{q}^{(r)}\big) +\frac{\kappa_{q}-\kappa_{q}^{(r)}}{\ln (2) \kappa_{q}^{(r)}} \Biggr)  \leq \upsilon_{q,2}, \label{36}
\end{align}
where $\kappa_{q}^{(r)}$ denotes the value of $\kappa_{q}$ in $r$-th iteration.

Finally, problem $\textbf{(P3)}$ is recast into the following equivalent form
\begin{subequations}\label{P4}
	\begin{align}
		\text{\textbf{(P4)}}	& \mathop {\max }\limits_{( \mathbf {W}_{c}, \mathbf {W}_{q}, r^c_q, \Psi^p_{q,x}, \boldsymbol{\beta}_{q}, \boldsymbol{\upsilon}_q, \kappa_q )} \    \sum\limits_{{\substack{q=1}}}^Q \Big(r^c_q + \Psi^p_{q,x}\Big), \label{37a}  \\
		s.t.\ \ & \eqref{29a}, \eqref{30a}, \eqref{30b}, \eqref{31}, \eqref{32}, \eqref{33}, \eqref{35b}, \eqref{36}, \label{37b}\\
		\ \ & \mathbf W_c \succeq 0, \ \mathbf W_q \succeq 0, \ \forall q\in \mathcal Q, \label{37c}\\ 
		\ \ & \operatorname{rank}(\mathbf W_c)=1, \ \operatorname{rank}(\mathbf W_q)=1,  \forall q\in \mathcal Q. \label{37d}
	\end{align}
\end{subequations}

It is worth noting that the non-convexity of problem \textbf{(P4)} arises from the rank-one constraints specified in \eqref{37d}. To tackle this difficulty, we reformulate \textbf{(P4)} as a convex semidefinite programming (SDP) problem by relaxing the rank-one constraints using the semidefinite relaxation (SDR) technique. Furthermore, if the obtained solutions $\mathbf{W}_c$ and $\mathbf{W}_q$ are of higher rank, we employ the Gaussian randomization method to construct rank-one solutions \cite{ni2021resource}.
  
\subsection{Optimization of $\mathbf \Theta_x$: Stage 2}
Given the transmit precoding vectors $\mathbf w_{c}$ and $\mathbf w_{q}$ obtained in Stage 1, the optimization problem for determining the passive beamforming matrix $\mathbf \Theta_x$ can be formulated as
\begin{subequations}\label{P5}
	\begin{align}
		\text{\textbf{(P5)}}	& \mathop {\max }\limits_{( r^c_q, \mathbf \Theta_x)} \    \sum\limits_{{\substack{q=1}}}^Q \Big(r^c_q + \widetilde{R}^p_{q,x}\Big), \label{38a}  \\
		s.t.\ \  & \eqref{27b}, \eqref{27c}, \eqref{15e}, \eqref{15f}. \label{38b} 
	\end{align}
\end{subequations}

Since \textbf{(P5)} is non-convex, we transform it into a tractable form by introducing a PSD matrix $\widehat{\mathbf {E}}_x = \hat{\mathbf{e}}_{x} \hat{\mathbf{e}}_{x}^H$ together with the constraints $\widehat{\mathbf{E}}_x \succeq \mathbf{0}$ and $\operatorname{rank}(\widetilde{\mathbf{E}}_x)=1$. Accordingly, \eqref{25} and \eqref{26} can be updated as
\begin{equation}
	\widehat{R}^c_{q,x}=\log _2\left(1+\frac{ \operatorname{tr}\Big(\widehat{\mathbf {E}}_x\widehat{\mathbf {F}}_{1}\Big)-\varepsilon_{qx}^2 \operatorname{tr}\Big(\mathbf{w}_c\mathbf{w}^H_c\Big)}
		{\operatorname{tr}\Big(\widehat{\mathbf {E}}_x\widehat{\mathbf {F}}_{2}\Big)+\varepsilon_{qx}^2 \operatorname{tr}\Big(\mathbf \Pi_c\Big)
			+ \tilde{\sigma}_q^2 }\right), \label{39}
\end{equation}
\begin{equation}
	\widehat{R}^{p}_{q,x}=\log _2\left(1+\frac{ \operatorname{tr}\Big(\widehat{\mathbf {E}}_x\widehat{\mathbf {A}}_{1}\Big)-\varepsilon_{qx}^2 \operatorname{tr}\Big(\mathbf{w}_q\mathbf{w}^H_q\Big) }
		{\operatorname{tr}\Big(\widehat{\mathbf {E}}_x\widehat{\mathbf {A}}_{2}\Big)+\varepsilon_{qx}^2 \operatorname{tr}\Big(\mathbf \Pi_p\Big)  + \tilde{\sigma}_q^2 }\right), \label{40}
\end{equation}
where $\widehat{\mathbf {F}}_{1}= \mathbf{H}_{d}\mathbf{w}_c\mathbf{w}^H_c\mathbf{H}_{d}^H \mathbf h_{qx}\mathbf h_{qx}^H$, $\mathbf \Pi_c= \mathbf \Pi_1+\mathbf \Pi_2$, $\widehat{\mathbf {F}}_{2}=\mathbf{H}_{d}\mathbf \Pi_c\mathbf{H}_{d}^H \mathbf h_{qx}\mathbf h_{qx}^H$, $\widehat{\mathbf {A}}_{1}= \mathbf{H}_{d}\mathbf{w}_q\mathbf{w}^H_q\mathbf{H}_{d}^H \mathbf h_{qx}\mathbf h_{qx}^H$, $\mathbf \Pi_p= \mathbf \Pi_2+\mathbf \Pi_3$, and $\widehat{\mathbf {A}}_{2}=\mathbf{H}_{d}\mathbf \Pi_p\mathbf{H}_{d}^H \mathbf h_{qx}\mathbf h_{qx}^H$. 

Consequently, problem \textbf{(P5)} can be equivalently expressed as
\begin{subequations}\label{P6}
	\begin{align}
		\text{\textbf{(P6)}}	& \mathop {\max }\limits_{( r^c_q, \widehat{\mathbf{E}}_x)} \    \sum\limits_{{\substack{q=1}}}^Q \Big(r^c_q + \widehat{R}^p_{q,x}\Big), \label{41a}  \\
		s.t.\ \ & r^c_q + \widehat{R}^p_{q,x}  \geq {R}_{min}, \forall q\in \mathcal Q, \label{41b} \\
		\ \ & \sum\limits_{{\substack{q=1}}}^Q  r^c_q  \leq \widehat{R}^c_{q,x}, \forall q\in \mathcal Q, \label{41c}\\
		 &\operatorname{diag}\big(\widehat{\mathbf{E}}_r\big)+\operatorname{diag}\big(\widehat{\mathbf{E}}_t\big)=\mathbf{1}^M, \label{41d}\\
		\ \ & \widehat{\mathbf{E}}_x \succeq \mathbf{0}, \forall x \in \{r,t\}, \label{41e}\\ 
		\ \ & \operatorname{rank}(\widehat{\mathbf{E}}_x)=1, \forall x \in \{r,t\}. \label{41f} 
	\end{align}
\end{subequations}

Next, to facilitate the convexity analysis of \textbf{(P6)}, we define a slack variable $\widehat{\Upsilon}^p_{q,x}$ as follows
\begin{align}
	& r^c_q + \widehat{R}^p_{q,x}  \geq \widehat{\Upsilon}^p_{q,x}, \forall q\in \mathcal Q. \label{42}
\end{align}
Further, Eq. \eqref{42} can be written as    
\begin{align}
	& r^c_q + \log_{2}\Biggl( \operatorname{tr}\Big(\widehat{\mathbf {E}}_x\widehat{\mathbf {A}}_{1}\Big)-\varepsilon_{qx}^2 \operatorname{tr}\Big(\mathbf{w}_q\mathbf{w}^H_q\Big) + \operatorname{tr}\Big(\widehat{\mathbf {E}}_x\widehat{\mathbf {A}}_{2}\Big)\nonumber \\
	&+\varepsilon_{qx}^2 \operatorname{tr}\Big(\mathbf \Pi_p\Big)  + \tilde{\sigma}_q^2  \Biggr) - \log_{2}\Biggl( \operatorname{tr}\Big(\widehat{\mathbf {E}}_x\widehat{\mathbf {A}}_{2}\Big)\nonumber +\varepsilon_{qx}^2 \operatorname{tr}\Big(\mathbf \Pi_p\Big) \nonumber\\
	& + \tilde{\sigma}_q^2  \Biggr)  \geq \widehat{\Upsilon}^p_{q,x}, \forall q\in \mathcal Q. \label{43}
\end{align}   

Given that \eqref{43} retains its non-convex nature, we utilize the SCA approach to linearize the expression as follows
\begin{align}
	& r^c_q + \log_{2}\Biggl( \operatorname{tr}\Big(\widehat{\mathbf {E}}_x\widehat{\mathbf {A}}_{1}\Big)-\varepsilon_{qx}^2 \operatorname{tr}\Big(\mathbf{w}_q\mathbf{w}^H_q\Big) + \operatorname{tr}\Big(\widehat{\mathbf {E}}_x\widehat{\mathbf {A}}_{2}\Big)\nonumber \\
	&+\varepsilon_{qx}^2 \operatorname{tr}\Big(\mathbf \Pi_p\Big)  + \tilde{\sigma}_q^2  \Biggr) - \widehat{\Omega}_1 \big(\widehat{\mathbf {E}}_x   \big)  \geq \widehat{\Upsilon}^p_{q,x}, \forall q\in \mathcal Q, \label{44}
\end{align}
 where $\widehat{\Omega}_1 \big(\widehat{\mathbf {E}}_x\big)$ denotes the first-order Taylor approximation given as follows
    \begin{align}
    	\widehat{\Omega}_1 \big(\widehat{\mathbf {E}}_x\big)&=\log_{2}\Biggl( \operatorname{tr}\Big(\widehat{\mathbf {E}}^{(r)}_x \widehat{\mathbf {A}}_{2}\Big) +\varepsilon_{qx}^2 \operatorname{tr}\Big(\mathbf \Pi_p\Big) + \tilde{\sigma}_q^2  \Biggr)  \nonumber \\
    	& \scalemath{0.92}{+ \mathrm{tr} \left(  \frac{\widehat{\mathbf {A}}_{2}\big(\widehat{\mathbf {E}}_x - \widehat{\mathbf {E}}^{(r)}_x\big)}{\left(\operatorname{tr}\Big(\widehat{\mathbf {E}}^{(r)}_x\widehat{\mathbf {A}}_{2}\Big) +\varepsilon_{qx}^2 \operatorname{tr}\Big(\mathbf \Pi_p\Big) + \tilde{\sigma}_q^2 \right)\ln 2 }  \right)}, \label{45}  
    \end{align}
with $\widehat{\mathbf {E}}^{(r)}_x$ represents the value of $\widehat{\mathbf {E}}_x$ obtained in the $r$-th iteration.

Further, to facilitate the convexity of \eqref{41c}, we can write as
\begin{align}
	&  \log_{2}\Biggl( \operatorname{tr}\Big(\widehat{\mathbf {E}}_x\widehat{\mathbf {F}}_{1}\Big)-\varepsilon_{qx}^2 \operatorname{tr}\Big(\mathbf{w}_c\mathbf{w}^H_c\Big) + \operatorname{tr}\Big(\widehat{\mathbf {E}}_x\widehat{\mathbf {F}}_{2}\Big)\nonumber\\
	&+\varepsilon_{qx}^2 \operatorname{tr}\Big(\mathbf \Pi_c\Big)
	+ \tilde{\sigma}_q^2  \Biggr) - \log_{2}\Biggl( \operatorname{tr}\Big(\widehat{\mathbf {E}}_x\widehat{\mathbf {F}}_{2}\Big)+\varepsilon_{qx}^2 \operatorname{tr}\Big(\mathbf \Pi_c\Big) \nonumber \\
	&+ \tilde{\sigma}_q^2  \Biggr)  \geq \sum\limits_{{\substack{q \in \mathcal Q}}}  r^c_q , \forall q\in \mathcal Q. \label{46}
\end{align} 

Since the left-hand side of \eqref{46} retains its non-convexity due to the second logarithmic component. Thus, we employ SCA as follows
\begin{align}
	&  \log_{2}\Biggl( \operatorname{tr}\Big(\widehat{\mathbf {E}}_x\widehat{\mathbf {F}}_{1}\Big)-\varepsilon_{qx}^2 \operatorname{tr}\Big(\mathbf{w}_c\mathbf{w}^H_c\Big) + \operatorname{tr}\Big(\widehat{\mathbf {E}}_x\widehat{\mathbf {F}}_{2}\Big)\nonumber\\
	&+\varepsilon_{qx}^2 \operatorname{tr}\Big(\mathbf \Pi_c\Big)
	+ \tilde{\sigma}_q^2  \Biggr) - \widehat{\Omega}_2 \big(\widehat{\mathbf {E}}_x   \big)  \geq \sum\limits_{{\substack{q \in \mathcal Q}}}  r^c_q , \forall q\in \mathcal Q, \label{47}
\end{align}    
where $\widehat{\Omega}_2 \big(\widehat{\mathbf {E}}_x\big)$ denotes the first-order Taylor approximation, given as
    \begin{align}
	\widehat{\Omega}_2 \big(\widehat{\mathbf {E}}_x\big)&=\log_{2}\Biggl( \operatorname{tr}\Big(\widehat{\mathbf {E}}^{(r)}_x \widehat{\mathbf {F}}_{2}\Big) +\varepsilon_{qx}^2 \operatorname{tr}\Big(\mathbf \Pi_c\Big) + \tilde{\sigma}_q^2  \Biggr)  \nonumber \\
	& \scalemath{0.92}{+ \mathrm{tr} \left(  \frac{\widehat{\mathbf {F}}_{2}\big(\widehat{\mathbf {E}}_x - \widehat{\mathbf {E}}^{(r)}_x\big)}{\left(\operatorname{tr}\Big(\widehat{\mathbf {E}}^{(r)}_x\widehat{\mathbf {F}}_{2}\Big) +\varepsilon_{qx}^2 \operatorname{tr}\Big(\mathbf \Pi_c\Big) + \tilde{\sigma}_q^2 \right)\ln 2 }  \right)}. \label{48}  
\end{align}

Finally, the optimization problem \textbf{(P6)} is recast into the following equivalent form
 \begin{subequations}\label{P7}
 	\begin{align}
 		\text{\textbf{(P7)}}	& \mathop {\max }\limits_{( r^c_q, \widehat{\Upsilon}^p_{q,x}, \widehat{\mathbf{E}}_x)} \    \sum\limits_{{\substack{q=1}}}^Q \widehat{\Upsilon}^p_{q,x}, \forall q\in \mathcal Q, \label{49a}  \\
 		s.t.\ \ & \widehat{\Upsilon}^p_{q,x}  \geq {R}_{min}, \forall q\in \mathcal Q, \label{49b} \\
 		&\operatorname{diag}\big(\widehat{\mathbf{E}}_r\big)+\operatorname{diag}\big(\widehat{\mathbf{E}}_t\big)=\mathbf{1}^M, \label{49c}\\
 		\ \ & \widehat{\mathbf{E}}_x \succeq \mathbf{0}, \forall x \in \{r,t\}, \label{49d}\\ 
 		\ \ & \operatorname{rank}(\widehat{\mathbf{E}}_x)=1, \forall x \in \{r,t\}, \label{49e} \\
 		\ \ & \eqref{44}, \eqref{47}. \label{49f}
 	\end{align}
 \end{subequations}
It is worth noting that problem \textbf{(P7)} is still non-convex and computationally challenging, primarily due to the rank-one constraint in \eqref{49e}. To address this issue, we reformulate \text{\textbf{(P7)}} as a convex SDP problem by relaxing the rank-one constraint using the SDR technique. Additionally, if $\operatorname{rank}(\widehat{\mathbf{E}}_x)\neq 1$, we employ the Gaussian randomization method to construct a rank-one solution.
\begin{algorithm}[t]
	\caption{Proposed Robust Optimization Framework}
	\begin{algorithmic}[1]
		\State \textbf{Initialization:} Initialize the system parameters, $ \hat{\mathbf W}_c$, $\hat{\mathbf W}_q$, $\hat{\mathbf \Theta}_x$, $\hat r_q^c$, $\hat \Psi^p_{q,x},  \boldsymbol{\hat \beta}_{q}, \boldsymbol{\hat \upsilon}_q, \hat \kappa_q, \hat{\Upsilon}^p_{q,x}, \hat{\mathbf{E}}_x$\\
		
		\noindent\textbf{Stage 1:} Optimization of transmit precoding vectors, $\mathbf W_c$, $\mathbf W_q$
		\While{convergence is not achieved and iteration count $\leq \hat{\tau}_{\max}$}
		\While{not converged and iteration count $\leq \hat{\mu}_1$}
		\State \parbox[t]{0.80\linewidth}{Compute $\mathbf W_c$, $\mathbf W_q$, and $r_q^c$ by solving the SDP problem \textbf{(P4)}}
		\If{$\text{rank}(\mathbf W_c)=1$ and $\text{rank}(\mathbf W_q)=1$}
		\State \parbox[t]{0.80\linewidth}{Obtain $\mathbf w_c$ and $\mathbf w_q$ using SVD of $\mathbf W_c$ and $\mathbf W_q$}
		\Else
		\State \parbox[t]{0.80\linewidth}{Apply Gaussian randomization method to construct rank-one solutions}\\
		\EndIf
		
		\State Update $\hat{\mathbf W}_c \leftarrow \mathbf W_c$ and $\hat{\mathbf W}_q \leftarrow \mathbf W_q$
		\EndWhile \\
		
		\noindent\textbf{Stage 2:} Optimization of transmission and reflection beamforming, $\mathbf \Theta_x$
		\While{not converged and iteration count $\leq \hat{\mu}_2$}
		\State \parbox[t]{0.85\linewidth}{Compute $\widehat{\mathbf{E}}_x$ and $r_q^c$ by solving the SDP problem \textbf{(P7)}}
		
		\If{$\text{rank}(\widehat{\mathbf{E}}_x=1)$}
		\State \parbox[t]{0.85\linewidth}{Obtain $\hat{\mathbf{e}}_{x}$ using SVD of $\widehat{\mathbf{E}}_x$}
		\Else
		\State \parbox[t]{0.80\linewidth}{Apply Gaussian randomization method to obtain rank-one solution}\\
		\EndIf
		
		\State Update $\hat{\mathbf{E}}_x \leftarrow \widehat{\mathbf{E}}_x$
		\EndWhile
		
		\EndWhile
		\State \Return $\mathbf w^{*}_c$, $\mathbf w^{*}_q$, $\mathbf \Theta^{*}_x$ 
	\end{algorithmic}
\end{algorithm}

\subsection{Proposed Robust Algorithm: Complexity and Convergence Analysis}    
1) \textit{Computational complexity:} In \textbf{Algorithm1}, we introduce a two-stage optimization strategy tailored for the considered STAR-RIS-enabled multi-user RSMA system. The main source of computational complexity lies in solving sub-problems \textbf{(P4)} and \textbf{(P7)}. In the first stage, the convex semi-definite programming problem \textbf{(P4)} is handled via the CVX toolbox employing the MOSEK solver, leading to a complexity of $\scalemath{0.85}{\mathcal{O}\bigl(\hat{\mu}_1 N^{3.5}\bigr)}$ \cite{wright1997primal,ben2001lectures}, where $\hat{\mu}_1$ represents the number of iterations required for Stage 1 to converge. In the second stage, sub-problem \textbf{(P7)} is similarly solved using CVX, which incurs a complexity of $\scalemath{0.85}{\mathcal{O}\bigl(\hat{\mu}_2 M^{3.5}\bigr)}$ \cite{wright1997primal,ben2001lectures}, with $\hat{\mu}_2$ denoting the iterations needed for Stage 2 convergence. Therefore, the total computational complexity of \textbf{Algorithm1} can be quantified as $\scalemath{0.85}{\mathcal{O}\Big[{\hat \tau}_{max}\big(\hat{\mu}_1 N^{3.5} + \hat{\mu}2 M^{3.5}\big)\Big]}$, where ${\hat \tau}_{max}$ denotes the total number of iterations required for the overall alternating optimization framework to converge.

\textit{2) Convergence Analysis:} Let $\mathbf w^{(m)}_c$, $\mathbf w^{(m)}_q$, and $\mathbf \Theta^{(m)}_x$ denote the solutions of the optimization variables at the $m$-th iteration. Accordingly, the objective function of the considered problem can be written as $\widehat R\big(\mathbf w^{(m)}_c, \mathbf w^{(m)}_q, \mathbf \Theta^{(m)}_x\big)$. In Stage 1, given a fixed $\mathbf \Theta^{(m)}_x$, the optimal transmit precoding vectors $\mathbf w^{*}_{c}$ and $\mathbf w^{*}_{q}$ are obtained by solving sub-problem \textbf{(P4)}. Consequently, we can have
\begin{align}
\widehat R\big(\mathbf w^{(m)}_c, \mathbf w^{(m)}_q, \mathbf \Theta^{(m)}_x\big) \leq \widehat R\big(\mathbf w^{(m+1)}_c, \mathbf w^{(m+1)}_q, \mathbf \Theta^{(m)}_x\big). \label{50}
\end{align}  

In \textbf{Stage 2}, the passive beamforming scattering matrix $\mathbf \Theta^{*}_x$ is computed for the given transmit precoding vectors $\mathbf w^{(m)}_c$ and $\mathbf w^{(m)}_q$. Therefore, we obtain
\begin{align}
	\widehat R\big(\mathbf w^{(m+1)}_c, \mathbf w^{(m+1)}_q, \mathbf \Theta^{(m)}_x\big) \leq \widehat R\big(\mathbf w^{(m+1)}_c, \mathbf w^{(m+1)}_q, \mathbf \Theta^{(m+1)}_x\big). \label{51}
\end{align}  

As a result, we obtain as
\begin{align}
	\widehat R\big(\mathbf w^{(m)}_c, \mathbf w^{(m)}_q, \mathbf \Theta^{(m)}_x\big) \leq \widehat R\big(\mathbf w^{(m+1)}_c, \mathbf w^{(m+1)}_q, \mathbf \Theta^{(m+1)}_x\big). \label{52}
\end{align}  

Consequently, the derived sequence of inequalities confirms that the objective function increases monotonically (or remains constant) with each iteration of \textbf{Algorithm 1}. This monotonic behavior ensures the convergence of \textbf{Algorithm 1}

   \begin{figure}[!t]
   	\centering
   	\includegraphics [width=0.48\textwidth]{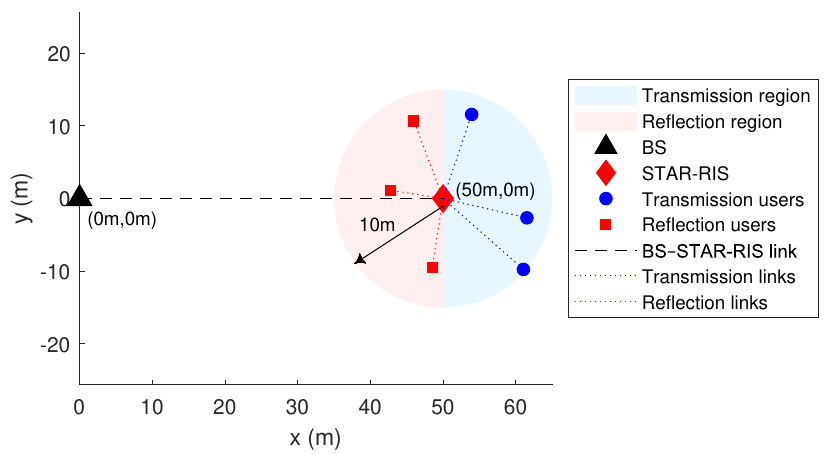}
   	\caption{Simulation environment.}
   	\label{f2}
   \end{figure}

\section{Simulation Results and Performance Analysis}
This section describes the simulation setup used to assess the performance of the proposed system considering HIs and channel estimation errors. As depicted in Fig.~\ref{f2}, the BS and the STAR-RIS are located at coordinates $(0~\text{m}, 0~\text{m})$ and $(50~\text{m}, 0~\text{m})$, respectively. In this setup, four users are considered, all of which are randomly distributed within a circular area of radius $10~\text{m}$ centered at the STAR-RIS, with two users located in the transmission region and the remaining two in the reflection region. Additionally, the channels $\mathbf{H}_d$ and $\mathbf{h}_{qx}$, $\forall q \in \mathcal{Q}$ and $\forall x \in \{r, t\}$, follow a Rician fading model, with their associated large-scale path losses expressed as $10^{-3} d^{-2}$ and $10^{-3} d^{-2.5}$, respectively, where $d$ represents the distance in meters \cite{yang2021reconfigurable}. Moreover, for both $\mathbf{H}_d$ and $\mathbf{h}_{qx}$ channels, the NLOS components are modeled as independent and identically distributed (i.i.d.) complex Gaussian random variables \cite{asif2025noma}, while the LOS components are characterized using steering vectors that incorporate both the angle of departure (AoD) and the angle of arrival (AoA) of each propagation path \cite{yang2021reconfigurable}. Unless otherwise specified, the key simulation parameters are set as follows: $\kappa_t=0.01$, $\kappa_r=0.01$, $M = 32$, $N = 8$, $\alpha_d = \alpha_{qx} = 10$, ${\sigma}_q^2 = -114$ dBm, bandwidth $= 1$ MHz, carrier frequency = $2.5$ GHz, $R_{\min} = 0.5$ Mbps, and $\varepsilon_{qx}^2 = \varepsilon^2 = \varepsilon \|\hat{\mathbf{H}}_{qx}\|$, where $\varepsilon \in [0, 1)$ \cite{zheng2023zero}.

Moreover, to evaluate the performance of the proposed technique, referred to as “STAR-Opt.,” three benchmark schemes are introduced. In particular,\textbf{ Benchmark-1} corresponds to solving problems (P4) and (P7) based on a traditional RIS configuration that serves only the users in reflection region. \textbf{Benchmark-2} solves (P4) to obtain the active precoding vectors with fixed values of passive beamforming vectors for both the transmission and reflection regions. Finally, \textbf{Benchmark-3} solves problems (P4) and (P7) by randomly allocating all optimization variables.

\begin{figure}[!t]
	\centering
	\includegraphics [width=0.45\textwidth]{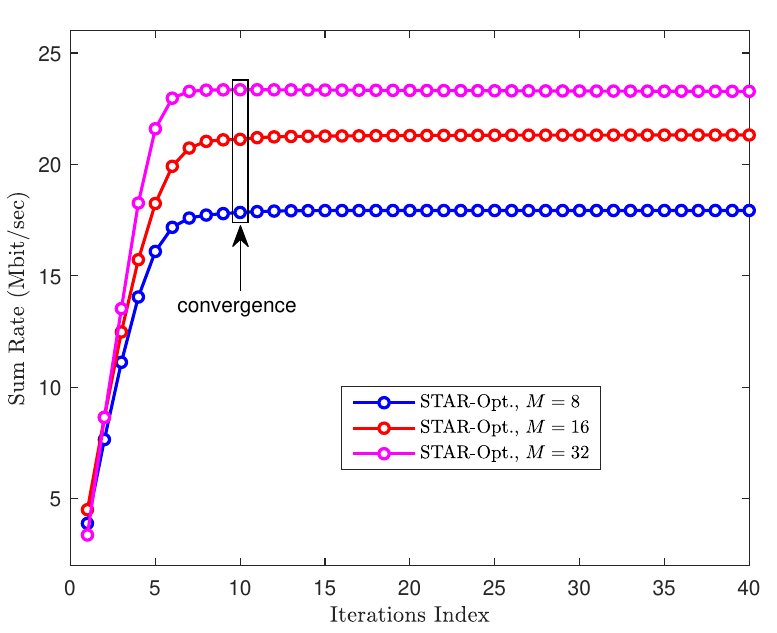}
	\caption{Convergence performance under different values of $M$.}
	\label{f3}
\end{figure} 

\begin{figure}[t]
	\centering
	\includegraphics [width=0.45\textwidth]{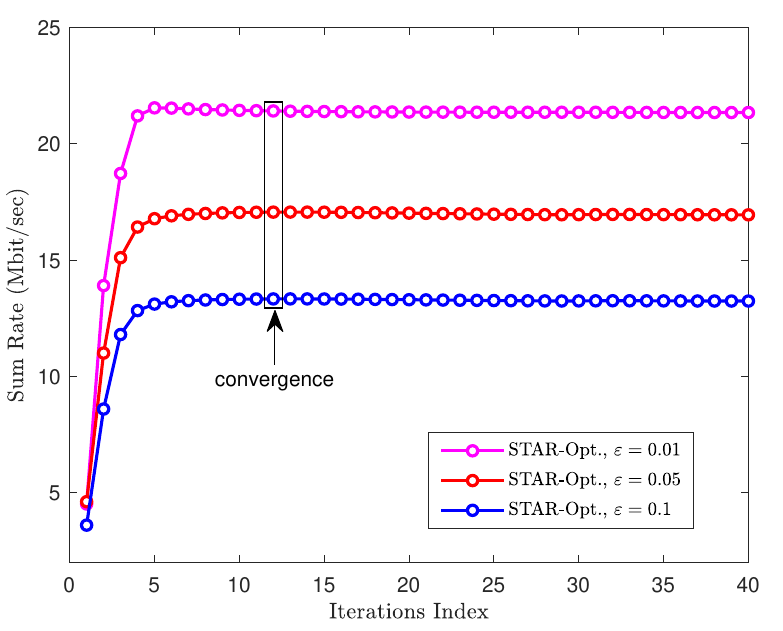}
	\caption{Convergence performance under different values of $\varepsilon$.}
	\label{f4}
\end{figure}

  \begin{figure}[!t]
	\centering
	\includegraphics [width=0.45\textwidth]{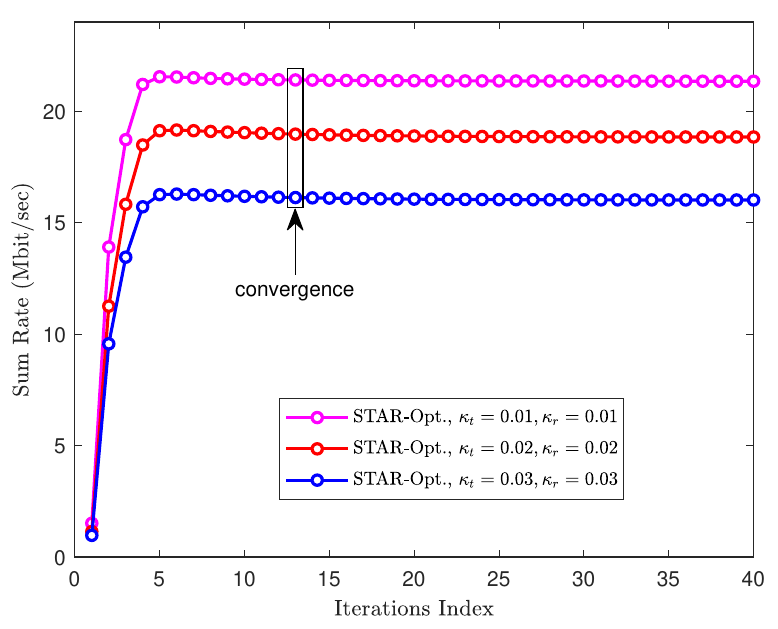}
	\caption{Convergence for different values of $\kappa_t$ and $\kappa_r$.}
	\label{f5}
\end{figure}

\begin{figure}[t]
	\centering
	\includegraphics [width=0.45\textwidth]{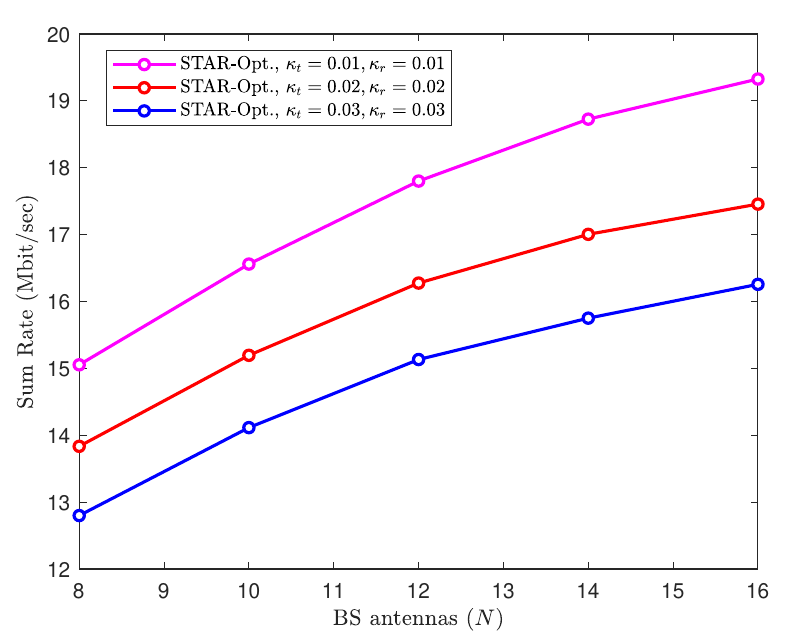}
	\caption{System's sum-rate for increasing values of $N$.}
	\label{f6}
\end{figure}

\begin{figure}[t]
\centering
\includegraphics [width=0.45\textwidth]{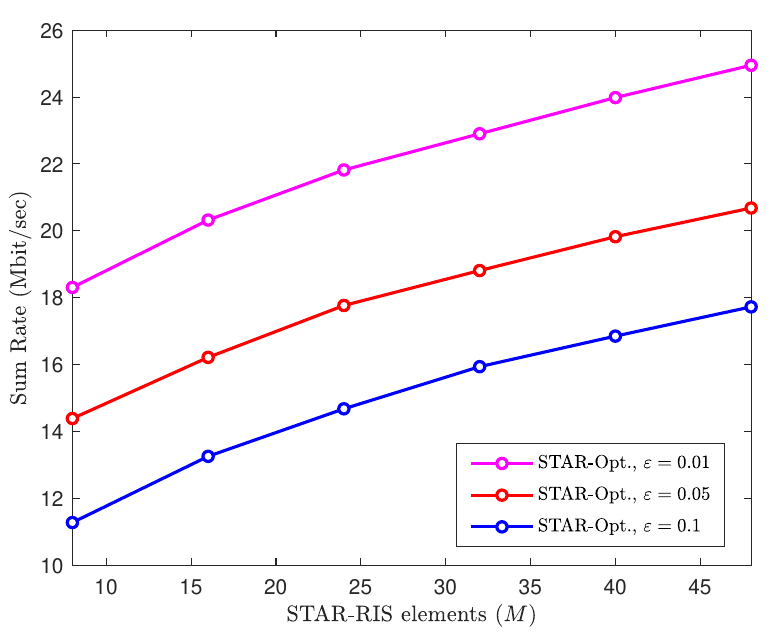}
\caption{Sum-rate performance for increasing values of $M$.}
\label{f7}
\end{figure}

\begin{figure}[t]
	\centering
	\includegraphics [width=0.45\textwidth]{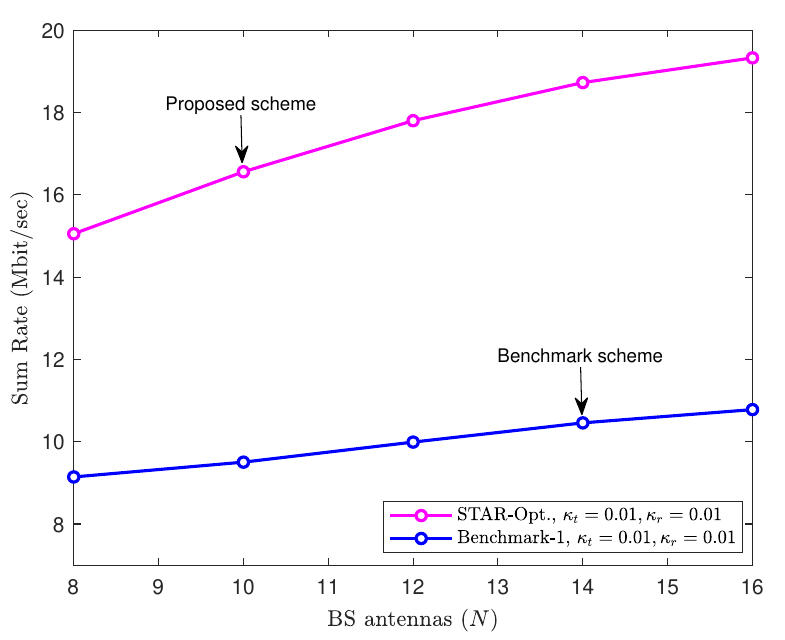}
	\caption{Comparative performance analysis under different values of $N$.}
	\label{f8}
\end{figure} 

\begin{figure}[t]
	\centering
	\includegraphics [width=0.45\textwidth]{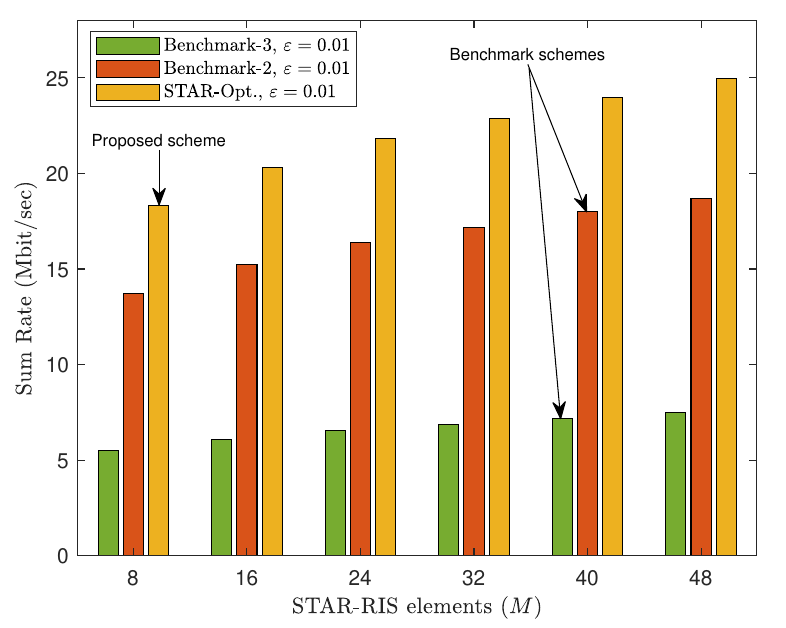}
	\caption{Comparative performance analysis under different values of $M$.}
	\label{f9}
\end{figure} 

 	\begin{figure}[t]
	\centering
	\includegraphics [width=0.45\textwidth]{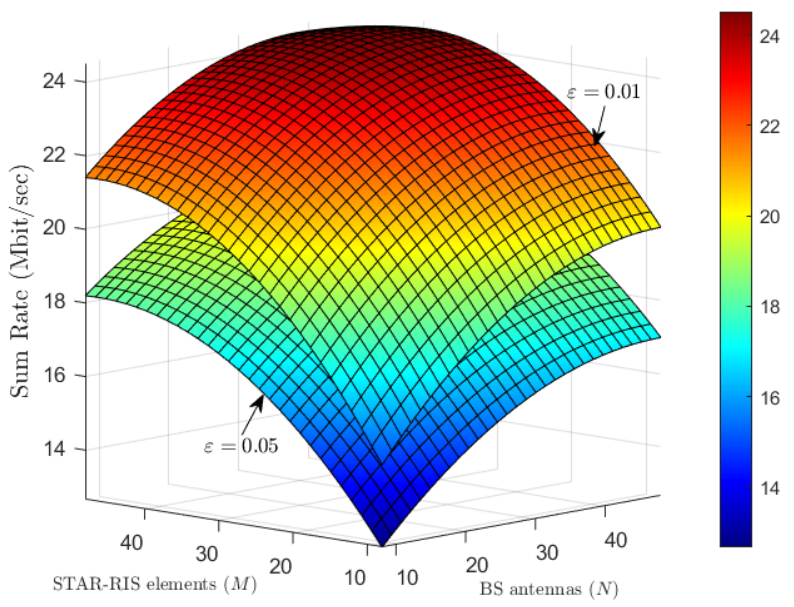}
	\caption{Joint impact of $N$ and $M$ under under different levels of channel uncertainty $\varepsilon$.}
	\label{f10}
\end{figure} 

\begin{figure}[t]
	\centering
	\includegraphics [width=0.45\textwidth]{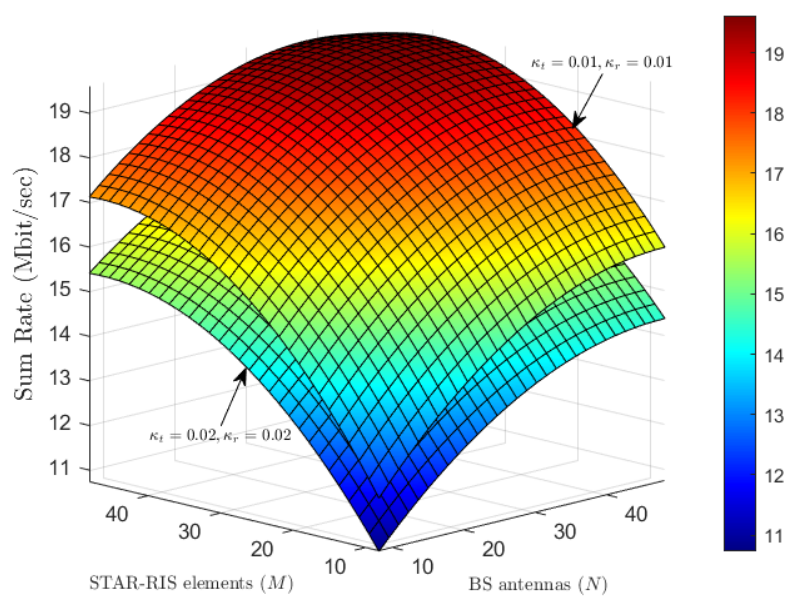}
	\caption{Joint impact of $N$ and $M$ under under different values of $\kappa_t$ and $\kappa_r$.}
	\label{f11}
\end{figure} 
As shown in Fig.~\ref{f3}, the convergence characteristics of the proposed algorithm are evaluated for different configurations of STAR-RIS elements. The system sum-rate increases with larger $M$, as more STAR-RIS elements enhance the beamforming gain. Fig.~\ref{f4} illustrates the convergence behavior under various channel uncertainty levels $\varepsilon$, where increasing $\varepsilon$ leads to performance degradation due to reduced accuracy in channel estimation. Similarly, Fig.~\ref{f5} presents the convergence characteristics for different transmitter and receiver hardware impairment factors, $\kappa_t$ and $\kappa_r$. The sum-rate declines with increasing $\kappa_t$ and $\kappa_r$, as more severe HIs significantly degrade the effective SINR. Consequently, the results presented in Figs.~\ref{f3}--\ref{f5} verify that the proposed algorithm achieves consistent and stable convergence within a practical number of iterations.

Next, Fig.~\ref{f6} depicts the influence of the number of BS antennas, $N$, on the achievable system sum-rate under various hardware impairment parameters $\kappa_t$ and $\kappa_r$. With larger $N$, the sum-rate increases because the expanded antenna array enhances the BS’s beamforming precision, thereby allowing more focused power delivery to the users. However, the sum-rate decreases with larger $\kappa_t$ and $\kappa_r$, as more severe HIs substantially degrade the effective SINR and, consequently, the overall system performance.

Moreover, Fig.~\ref{7} depicts the influence of the number of STAR-RIS elements, $M$, on the system sum-rate for various channel uncertainty levels $\varepsilon$. A clear improvement in sum-rate is observed as $M$ increases. This performance gain is attributed to the enhanced effective channel gain and the additional spatial degrees of freedom provided by a larger STAR-RIS, which collectively strengthen its passive beamforming capability. Additionally, it is noteworthy that the system sum-rate decreases with higher values of $\varepsilon$, since larger channel uncertainty leads to performance degradation caused by reduced channel estimation accuracy.

Subsequently, Fig.~\ref{8} compares the performance of the proposed approach, STAR-Opt., with the benchmark scheme as the number of BS antennas increases. The numerical results confirm that, for any number of BS antennas $N$, the proposed strategy yields higher sum-rate performance than the traditional RIS-based approach. This improvement arises because, unlike the conventional RIS that supports only half-space coverage, the STAR-RIS enables full-space ($360^\circ$) coverage by simultaneously serving users in both the transmission and reflection regions. Additionally, Fig.~\ref{9} compares the performance of the proposed STAR-Opt. scheme with Benchmark-2 and Benchmark-3 under varying numbers of STAR-RIS elements, $M$. The numerical results demonstrate that, for all tested values of $M$, the proposed method outperforms both benchmark schemes by delivering superior sum-rate performance.

Finally, Figs.~\ref{10} and \ref{11} illustrate the joint impact of the number of STAR-RIS elements, $M$, and BS antennas, $N$, on the system sum-rate under varying levels of channel uncertainty ($\varepsilon$) and hardware impairment parameters ($\kappa_t$ and $\kappa_r$), respectively. The numerical results reveal that the sum-rate is more sensitive to variations in $M$ than in $N$. This occurs because, in the absence of a direct communication link, the STAR-RIS acts as the primary propagation path. Thus, each additional STAR-RIS element enhances both the energy captured from the BS and the coherent re-radiation toward users in the transmission and reflection regions, thereby accelerating the growth of the effective channel gain with $M$. In contrast, increasing $N$ mainly strengthens the active beamforming capability at the BS, producing a relatively moderate improvement due to the fixed total transmit power being distributed among antennas. Consequently, the increase in STAR-RIS elements provides a more pronounced enhancement in the overall sum-rate compared to the increase in BS antennas.

\section{Conclusion}
This paper presented a robust joint optimization framework for a STAR-RIS-assisted multi-user RSMA downlink system under transceiver HIs and imperfect CSI. By integrating the flexibility of RSMA with the dual functionality of STAR-RIS, the proposed scheme enhances both spectral and energy efficiency, ensuring reliable connectivity even when the direct communication link is obstructed. The original highly non-convex sum-rate maximization problem was effectively addressed through an alternating optimization framework, wherein the transmit precoding and passive beamforming matrices were iteratively optimized by leveraging SCA, SDP, and SDR-based convex reformulations, followed by Gaussian randomization for rank-one recovery. Numerical simulations validated the superiority of the proposed design, demonstrating substantial performance gains over benchmark schemes and confirming its robustness and fast convergence. 


\bibliographystyle{IEEEtran}
\bibliography{Ref}
\vskip -2\baselineskip plus -2fil

\end{document}